\documentclass[a4paper, notoc]{JHEP3}
\usepackage{graphicx}
\usepackage{amsfonts}
\usepackage{amssymb}
\usepackage{epsfig}
\usepackage{cite}
\usepackage{textcomp}
\usepackage{graphicx,psfrag, subfigure}

 \newlength{\wth}
 \newcommand{\twographs}[2]{%
 \unitlength=1.1in
 \begin{picture}(5.8,2.6)
 \put(0,0){\epsfig{file=#1.eps, width=0.75\wth}}
 \put(2.5,0){\epsfig{file=#2.eps, width=0.75\wth}}
 \end{picture}
}

 \newcommand{\eqref}[1]{{(\ref{#1})}}

\newcommand{\be}{\begin{equation}}
\newcommand{\ee}{\end{equation}}
\newcommand{\bea}{\begin{eqnarray}}
\newcommand{\eea}{\end{eqnarray}}
\newcommand{\nn}{\nonumber}

\newcommand{\ti}{\times}
\newcommand{\half}{\frac{1}{2}}
\newcommand{\mc}{\mathcal}

\newcommand{\beqa}{\begin{eqnarray}}
\newcommand{\eeqa}{\end{eqnarray}}

\newcommand{\xx}{\xi_X}
\newcommand{\xa}{\xi_a}

 \title{The Cosmophenomenology of Axionic Dark Radiation}
\author{Joseph P. Conlon, M.C.~David Marsh
 \\ Rudolf Peierls Center for Theoretical Physics, 1 Keble Road \\
 Oxford OX1 3NP, UK \\ Email: \email{j.conlon1@physics.ox.ac.uk, david.marsh1@physics.ox.ac.uk}}

\abstract{Relativistic axions are good candidates for the dark radiation for which there are mounting observational hints.
 The primordial decays of heavy fields produce axions which are ultra-energetic compared to thermalised matter
 and inelastic axion-matter scattering can occur with $E_{CoM} \gg T_{\gamma}$, thus accessing
 many interesting processes which are otherwise kinematically forbidden in standard
cosmology.
Axion-photon scattering into quarks and leptons during BBN
affects the light element abundances, and bounds on  overproduction of $^4$He constrain a combination of the axion decay constant and the reheating temperature.
For supersymmetric models, axion scattering into
visible sector superpartners can give direct non-thermal production of dark
matter at $T_{\gamma} \ll T_{freezeout}$. Most axions --- or any other dark radiation candidate
from modulus decay --- still linger today 
as a Cosmic Axion Background with $E_{axion} \sim \mc{O}(100)~ \hbox{eV}$,
 and a flux of $\sim 10^6 ~\hbox{cm}^{-2} \hbox{s}^{-1}$.
}

\begin{document}
\tableofcontents
\section{Introduction}

Conventional cosmology begins with the Hot Big Bang. During this period, the energy density of the universe lies
entirely in thermalised Standard Model degrees of freedom, and the only particles with energies $E \gg T$ are non-relativistic.

However there are mounting hints that this standard picture of the early universe is not fully correct. In particular, there may exist
\emph{dark radiation}, i.e.~non-Standard Model particles that are relativistic at the times of Big Bang Nucleosynthesis (BBN) and at the formation of the Cosmic Microwave Background (CMB).
Including the HST measurement of the Hubble constant \cite{11032976}, the latest observational values from WMAP, ACT, SPT and Planck are $N_{eff} = 3.84 \pm 0.40$ (WMAP9, \cite{12125226}), $3.71 \pm 0.35$ (SPT, \cite{12126267}), $3.50 \pm 0.42$ (ACT, \cite{13010824}),  and $3.62 \pm 0.25$ (Planck, \cite{13035076}).\footnote{Without including direct measurements of $H_0$, the determinations using only CMB and BAO data are
$N_{eff} = 3.55 \pm 0.60$ (WMAP9), $2.87 \pm 0.60$ (ACT), $3.50 \pm 0.47$ (SPT) and $3.30 \pm 0.27$ (Planck). Note that there is some overlap
in the data sets in all cases and so the values should not be taken as independent. Also note that the straight $\Lambda$CDM fit to Planck
gives an inferred value for $H_0$ of $(67.3 \pm 1.2) \, \hbox{km s}^{-1} \hbox{ Mpc}^{-1}$ which is low by $2 \div 3 \sigma$ compared to
recent direct measurements \cite{11032976, 12083281, 12086010}.}
 While by no means conclusive, these hint at a non-Standard Model value of $N_{eff} > 3.046$, and the consequent existence
of dark radiation.

Dark radiation must involve a particle that is still relativistic at CMB formation at
$t \sim 3 \ti 10^5 \hbox{ years}$, which suggests that it is extremely light. 
One outstanding example of naturally light particles are axions --- either the QCD axion
or more generally an axion-like particle. Relativistic axions can also easily be produced non-thermally through the primordial decays of heavy
moduli or saxions.

There is a large literature on the cosmology of axions and a recent general review is \cite{12105081}.
There is also a growing literature on dark radiation, for example
\cite{ hepph0703034,10084528, 10105693, 10113501, 11074319, 11110605, 11111336, 12014816, 12060109, 12070497, 12072771, 12082496, 12082951,
12083562, 12083563, 12124160, 13017428, 13021486, 13022143, 13022516, 13030143, 13035379, 13036270}. Within these, \cite{12124160} gives a general review of dark radiation from particle decay, and
string models of axionic dark radiation directly relevant to this work are \cite{12083562, 12083563} (also see \cite{12072771}).
The role of relativistic \emph{thermal} axions in dark radiation has been considered in \cite{10084528, 12082951, 13022143}.

However what this literature all tends to assume is that, once established as a dark radiation candidate,
axions become non-dynamical.
The justification for this is that axion scattering is suppressed by $f_a^{-1}$, giving $\Gamma/H \ll 1$ and an axion population which is far from thermal equilibrium.
In particular, the effects of their scattering off the ambient thermal plasma are neglected.

In this paper we show that there is much interesting physics from the interactions of dark radiation axions with the thermal plasma.
In particular, it allows for a violation of some familiar properties of early universe cosmology. For example, in Standard Cosmology, there are no relativistic particles with energies $E \gg T$ since
the energy spectrum of particles is set by the Boltzmann
distribution. 
The ultimate abundance of a species, $Y = n/s$,
is determined either via the freeze out of stable
states (as in WIMP dark matter), or by a conserved quantum number (as in the case of baryon number, where $Y_B$ is far in excess
of the value given by thermal decoupling). Furthermore, any processes with a center of mass energy
$E \gg T$ are inaccessible: the thermal bath is simply incapable of generating collisions with such a high centre of mass energy
and they are suppressed by $e^{-E/T} \lll 1$.

With axionic dark radiation, these properties  no longer need hold.
Highly energetic axions may arise from very simple scenarios of the early universe in which
long-lived heavy particles of mass $m_{\Phi}$ reheat the Standard Model degrees of freedom to $T_{reheat} \ll m_{\Phi}$, while also decaying to dark radiation axions with $E_{axion} = m_{\Phi}/2$. The subsequent interaction of these
axions with the thermalised plasma allows processes at $E_{CoM} \sim \sqrt{T_{reheat} m_{\Phi}} \gg T_{reheat}$.

This scenario  is very well motivated: 
the existence of at least one axion (the QCD axion) has strong support from the non-observation
of a neutron electric dipole moment, and string theory models in general
can have many `axion-like particles' (hence, axions) which may not couple to QCD but still enjoy an approximate Peccei-Quinn shift symmetry. This symmetry forbids perturbative mass terms for the axion, making it naturally light. Furthermore axions are very weakly coupled to the visible sector --- for the QCD axion, experimental bounds require $10^9 \lesssim f_{PQ}/\hbox{GeV}$ ---
and require large reheating temperatures to thermalise. For example,  with $f_{PQ} = 10^{10} \hbox{GeV}$, the QCD
axion thermalises at $T \gtrsim 2 \ti 10^7~ \hbox{GeV}$ \cite{12082951}.


The existence of a long-lived massive particle species $\Phi$ which reheats the universe through its decay is similarly well-motivated.
 The relative scaling of matter ($\rho \sim a^{-3}$)
and radiation ($\rho \sim a^{-4}$) implies that, after inflation, the energy density of the universe tends to
become dominated by non-relativistic matter particles with the longest lifetimes.
As lifetime is inversely related to interaction strength,
these are naturally particles with interactions suppressed by $M_{Pl}^{-1}$ --- in the context of string compactifications these are
the ubiquitous moduli.
Particles with Planck-suppressed couplings
have a characteristic decay rate and reheat temperature
$$
\Gamma \sim \frac{1}{8 \pi} \frac{m_{\Phi}^3}{M_{Pl}^2}, \qquad T_{SM, reheat} \sim \frac{m_{\Phi}^{3/2}}{M_{Pl}^{\half}} = 0.6 \, \hbox{GeV} \left( \frac{m_{\Phi}}{10^6 \hbox{GeV}} \right)^{3/2},
$$
and will come to dominate the energy density of the universe and drive reheating. The existence of such particles
is generic in string theory or any extra dimensional theory, where they arise from the higher-dimensional modes of the graviton.

However, gravity induces very general couplings and there is no \emph{a priori} reason for $\Phi$ to couple only to the Standard Model.
On the contrary, in general the field $\Phi$ may
also decay into
any light axions that exist, and such axions 
contribute to the
`dark radiation' of the universe. From the decay $\Phi \rightarrow a a$, the initial energy of such axions is
$$
E_{axion} = \frac{m_{\Phi}}{2} \sim T_{reheat} \left( \frac{M_{Pl}}{m_{\Phi}} \right)^{\half} \gg T_{reheat} \, .
$$
Explicit models for such decays are discussed in \cite{hepph0703034, 12014816, 12083562, 12083563}.
As axions are very weakly coupled, they do \emph{not} thermalise after their original production,
and their original energy is only diluted through the redshift due to the expansion of the universe. In particular,
the axions remain highly energetic compared to the ambient photon plasma, by a factor $\left(M_{Pl}/m_{\Phi} \right)^{\half}$.

Although we have considered the particular case of Planck-coupled moduli above, the same picture will occur
whenever reheating is driven by the decays of long-lived massive states (provided they have a decay mode to axions).
Other examples could be topological field configurations, string winding states, Q-balls etc.
The key feature is simply that $m_{\Phi} \gg T_{reheat} \sim (H^2 M_{Pl}^2)^{1/4} \sim (\Gamma^2 M_{Pl}^2)^{1/4}$. In this case
the energy of hidden axions produced via direct 2-body decays of $\Phi$ is again very much greater than the particle energies in the
Standard Model thermal plasma.

We also note here, that while we focus on axions in this paper, many of the considerations also apply to other dark radiation candidates
such as hidden photons. The important point is only that the particles are produced from modulus decay, giving them a highly energetic
spectrum compared to that of the thermalised plasma.

The fact that the great majority of axions never interact with the ambient photon plasma does not mean no interactions
occur. Although such interactions may be rare, they are important because the centre of mass energy of
the interaction is far higher than can be achieved by purely thermal processes. Axion-photon scattering is then the
only way for some interactions to occur, as they are kinematically inaccessible within standard cosmology.

In this paper we study the effects of the scattering of highly energetic dark radiation axions off the ambient
Standard Model plasma. We focus our discussion on two processes:
\begin{enumerate}
\item
Inelastic scattering of axions off thermal photons into Standard Model degrees of freedom: $a + \gamma \to q_i \bar q_i$, where $q_i$ can be either a quark or a lepton. These processes can be important during and after Big Bang nucleosynthesis,  and may be constrained by their effects on the light element abundances.
\item
The non-thermal production of dark matter, after the conventional freeze-out temperature, by the scattering
of axions off the thermal plasma and into sparticles, $a + \gamma/g \to \tilde{q} \tilde{q}^{*}$.
\end{enumerate}

The paper is structured as follows. Section \ref{sec:gen} reviews the general formalism of moduli decay to give both reheating
and dark radiation. Section \ref{sec:scattering} discusses axion scattering and how it can give rise  to energy injections during BBN. 
 Section \ref{sec:dm} describes the generation of dark matter from scattering to supersymmetric particles, and in section \ref{sec:today}, we estimate the present day relic flux of relativistic axions.  The conclusions summarise and describe possible extensions of this work.

\section{Generalities of Moduli Decay} \label{sec:gen}

Throughout this paper, we assume that post-inflationary reheating is driven by the decay of a heavy modulus of mass $m_{\Phi}$ and decay rate $\Gamma$.
For a Planck-coupled modulus, $\Gamma$ and $m_{\Phi}$ are related by
\be
\Gamma = \frac{1}{4 \pi} \frac{m_{\Phi}^3}{(M_{Pl}/\kappa)^2} \, ,
\ee
where $\kappa$ is an $\mc{O}(1)$ constant and $M_{Pl}$ denotes the reduced Planck mass.  The modulus lifetime is then
\be
\tau = \frac{4 \pi}{\kappa^2} \frac{M_{Pl}^2}{m_{\Phi}^3} \, .
\ee
Such a modulus is long-lived and we assume that prior to its decay it dominates the energy density of the universe.
In the bulk of this paper we will use  the simultaneous decay approximation in which all moduli are assumed to decay precisely at time $\tau$,
although for determining the present-day axion energy spectrum in section \ref{sec:today} we will be more precise.

Prior to decay, the universe is matter dominated and $a \sim t^{2/3}$. As $H_{decay} = \frac{2}{3t}$ during matter domination, we have
\be
H_{decay} = \frac{\kappa^2}{6 \pi} \frac{m_{\Phi}^3}{M_{Pl}^2}.
\ee
We assume the modulus decays with a fraction $B_a$ to an axion (constituting dark radiation) and
a fraction $1-B_a$ to the visible Standard Model sector. We will assume
that the dark radiation component is single species and thus only consists of this axion. The initial axion energy density is then
\be
\rho_{0,axion} = B_a \ti 3 H_{decay}^2 M_{Pl}^2 \, ,
\ee
and the initial Standard Model energy density is given by,
\be
\rho_{0,SM} = (1-B_a) \ti 3 H_{decay}^2 M_{Pl}^2 \, .
\ee
As $\rho_{SM} = \frac{\pi^2}{30} g_{*}(T) T^4$, with $g_{*}(T)$ the effective number of species, we have
\be
T_{reheat} = \left( \frac{90}{\pi^2 g_{*}(T_{reheat} )} (1-B_a) H_{decay}^2 M_{Pl}^2 \right)^{1/4} = \kappa \left( \frac{5(1-B_a)}{2 \pi^4 g_{\star}(T_{reheat} )}\right)^{1/4} \frac{m^{3/2}_{\Phi}}{M^{1/2}_{Pl}} \, . \label{eq:Trh}
\ee
The initial axion number density at time of reheating is
\be
n_{axion,0} = B_a \ti \frac{3 H_{decay}^2 M_{Pl}^2 }{E_a} = \frac{B_a \kappa^4}{6 \pi^2} \frac{m_{\Phi}^5}{M_{Pl}^2} \, . \label{eq:n0}
\ee
In the non-interacting limit, the energy density of the relativistic axions evolves with the size of the universe as
\be
\rho_{axion}(R) = \rho_{0,axion} \left( \frac{R_0}{R} \right)^4,
\ee
whereas the energy density of the Standard Model plasma evolves as
\be
\rho_{SM}(R) = \rho_{0,SM} \left( \frac{g_{*,0}}{g_{*}} \right)^{1/3} \left( \frac{R_0}{R} \right)^4,
\ee
such that the comoving entropy $S \simeq g_{*} T^3 R^3$  is conserved during the expansion of the universe. As a result the
fraction of energy density in Standard Model degrees of freedom gradually increases as the universe expands and $g_{*}$ decreases.
The hidden sector branching ratio $B_a$ relates to the experimental observable $\Delta N_{eff} \equiv N_{eff} - 3.046$ as (e.g. see \cite{Choi:1996vz, 12083562}),
\be
\Delta N_{eff} = \frac{43}{7} \frac{B_a}{1-B_a} \left( \frac{g_{*}(T_{\nu \, decoupling})}{g_*(T_{reheat})} \right)^{1/3}.
\ee

After modulus decay, the Standard Model sector rapidly attains thermal equilibrium and
the number densities and energy distribution of particles in it are set by the Boltzmann distribution.
The axion decay mode $\Phi \to a a$ however produces monoenergetic particles with an energy $E_a = m_{\Phi}/2$.
This energy redshifts as $R^{-1}$ but does not otherwise change.\footnote{The monoenergetic spectrum is a feature of
the instantaneous decay approximation. In practice the decays will occur at different times and the spectrum will be
smeared by redshifting.}

As a numerical example we will use throughout this paper, if $m_{\Phi} = 5 \ti 10^6 \, \hbox{GeV}$ and $\kappa = 1$, then a hidden sector
branching ratio of $B_a = 0.145$ corresponds to a reheat temperature of $T_{reheat} = 1.00 \, \hbox{GeV}$. This branching ratio is
chosen to correspond to the Planck + HST central value of
 $N_{eff} = 3.62$, using $g_*(T_{reheat}) = 61.75$ and $g_*(T_{\nu \, decoupling}) = 10.75$. The choice of a decaying modulus mass in the region $m_{\Phi} = 10^6 \div 10^7 \hbox{GeV}$ is motivated by the appearance of this scale in various string scenarios \cite{hepph0503216, 08040863, 09063297}.

\begin{figure}[]
\begin{center}
\includegraphics[width=.85 \textwidth]{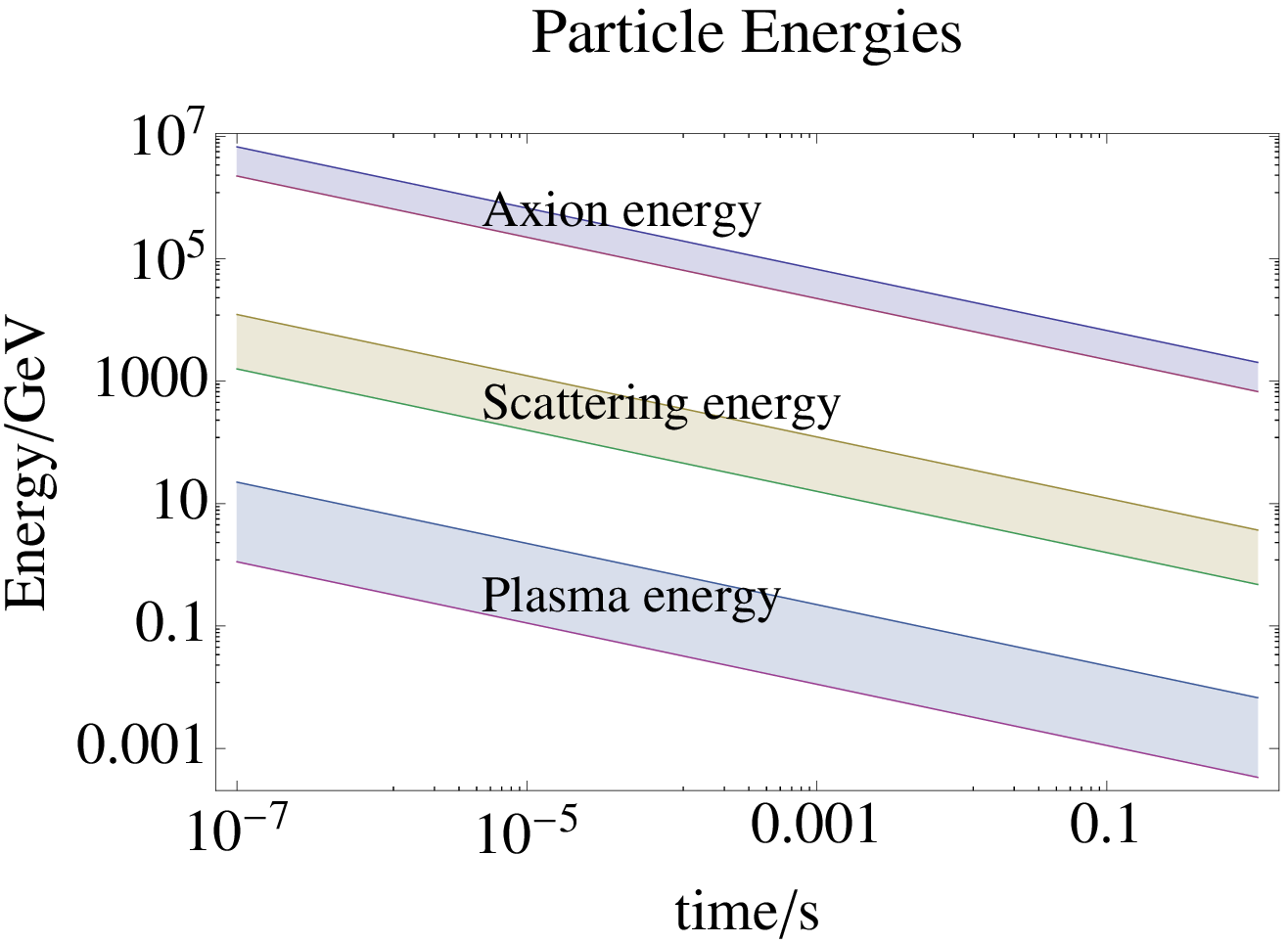}
\end{center}
\vspace{-5ex}
\caption{The three distinct energy scales and their evolution with time: the energy of the relativistic axions, the energy of the thermal Standard Model plasma, and the centre of mass energy for scattering between the axions and the thermal plasma. These can be separated by several orders of magnitude. 
}
\label{fig:AllEnergies}
\end{figure}

In this paper, our main interest will lie in the scattering of the relativistic axions off the thermal Standard Model plasma. In figure \ref{fig:AllEnergies} we
illustrate the approximate distribution of energies as the universe evolves, and how they differ from standard cosmology.

\section{Axion-Photon Scattering Constraints from BBN} \label{sec:scattering}

In this section we compute the inelastic scattering rate for an axion off thermal photons and determine the resulting constraints
due to energy injections during BBN.
We consider the process,
\be
a + \gamma \to q + \bar{q} \, , \label{eq:proc}
\ee
for the fermion/antifermion pair $q \bar q$ as placeholder for 
processes including axion scattering into quarks, $a + \gamma \to q_i + \bar q_i$, or any leptons, $a + \gamma \to l_i^{-} + l_i^{+}$.

\subsection{Calculation}

\begin{figure}
\begin{center}
\includegraphics[width=14cm]{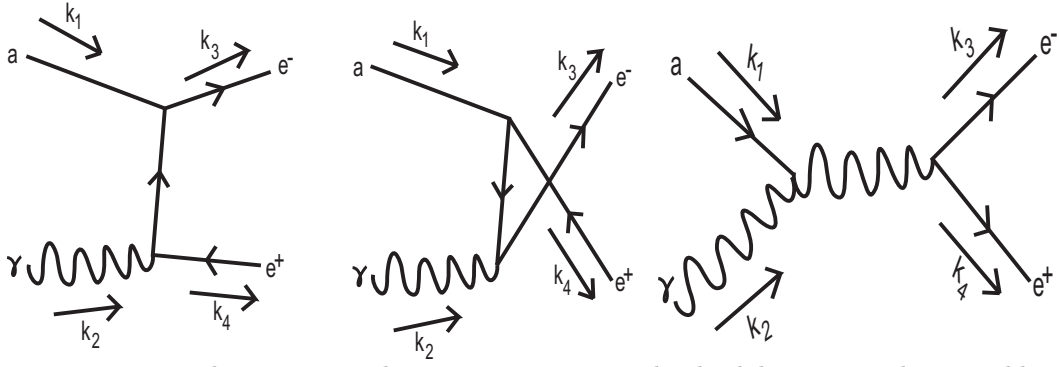}
\end{center}
\vspace{-5ex}
\caption{Feynman diagrams contributing to $a + \gamma \to q\bar q$. The third diagram involves an additional factor of $\alpha_{EM}$
and so we neglect it.}
\label{Figure}
\label{fig:Fdiag}
\end{figure}

The potential Lagrangian interaction terms are
\bea
\hbox{Axion-fermion-fermion:} & \qquad & c_q \frac{m_q}{f_a} a \bar{\psi} \gamma^5 \psi \, , \nn \\
\hbox{Photon-fermion-fermion:} & \qquad & q_q A_{\mu} \bar{\psi} \gamma^{\mu} \psi \, , \nn \\
\hbox{Axion-photon-photon:} & \qquad & c_{\gamma} \frac{\alpha_{em}}{8 \pi}  \frac{a}{f_a} \epsilon_{\mu \nu \lambda \rho}
F^{\mu \nu} F^{\lambda \rho} \equiv g_{a \gamma \gamma} a {\bf{E}} \cdot {\bf{B}} \, , \label{eq:couplings}
\eea
where $c_q$ and $c_{\gamma}$ are model-dependent constants. We shall henceforth take $c_q = 1$.
The process \eqref{eq:proc} proceeds via the Feynman diagrams shown in figure \ref{fig:Fdiag}.
As the third diagram involves the $a\gamma \gamma$ vertex which is suppressed by an additional factor of $\frac{\alpha_{em}}{4 \pi}$, we shall not
consider it further here.\footnote{The presence of $\alpha_{EM}$ can be understood
by starting with the `natural' form of the gauge kinetic action, $\frac{1}{4 g_{EM}^2} F_{\mu \nu} F^{\mu \nu} + \frac{a}{f_a} F_{\mu \nu} \tilde{F}^{\mu \nu}$, and then canonically normalising the kinetic terms.}

We aim to compute the axion scattering rate,
\be
\Gamma_{a \gamma \to q \bar q} = \Gamma = \langle n \sigma v \rangle \, ,
\ee
by first computing $\sigma v$, averaged over final state momenta and initial directions, for an axion of energy $E_a$ scattering on a photon of energy $E_{\gamma}$, and then subsequently performing the thermal averaging over $E_{\gamma}$.

The axion and photon 4-momenta are
\bea
k_{axion} & = & (E_a, E_a, 0, 0), \nn \\
k_{photon} & = & (E_{\gamma}, E_{\gamma} \cos \theta, E_{\gamma} \sin \theta, 0). \nn
\eea
where we have assumed  $m_a \ll E_a$. The centre of mass energy is then given by
$E_{CoM}^2 = 2 k_{axion} \cdot k_{photon} 
 =  2 E_a E_{\gamma} (1 - x)$
with $x = \cos \theta$.

Denoting the amplitudes for the first two diagrams in figure \ref{fig:Fdiag} by $\mc{M}_A$ and $\mc{M}_B$ respectively, we have
\be
\frac{1}{2} \sum_{\rm spins}\vert \mc{M}_{A} + \mc{M}_B \vert^2 = \frac{2 q_q^2 m_q^2}{f_a^2} \left( \frac{u}{t} + \frac{t}{u} +2 \right),
\ee
where the first term arises from $\vert \mc{M}_A \vert^2$, the second from $\vert \mc{M}_B \vert^2$, and the third from the cross term.
Here $t = k_1 \cdot k_3$ and $u = k_1 \cdot k_4$. Integrating over the $q \bar q$ phase space, we get
\be
\sigma = \frac{1}{4 E_a E_{\gamma} \vert v_a - v_{\gamma} \vert } \frac{q_q^2 m_q^2}{2 \pi f_a^2}
 \ln \left( \frac{E^{'} + p^{'}}{E^{'} - p^{'}} \right)\, . \label{eq:sigma}
\ee
Here $E^{'} = E_{CoM}/2$ 
 is the energy of the fermion in the centre of mass frame and
$p^{'} = \sqrt{(E^{'})^2 - m_q^2}$ is the momentum of the fermion in the centre of mass frame.

We can average $\sigma v$ over the initial photon direction by performing the $x$-integral, $\half \int_{-1}^{1- \lambda} dx$, where $\lambda = \frac{2m_q^2}{E_a E_{\gamma}}$ accounts for the threshold energy for $q \bar q$ pair production.
The relevant integral is
\bea
\langle \sigma v \rangle &=&   \frac{q_q^2 \lambda}{16 \pi f_a^2} \int_{-1}^{1-\lambda} dx  \ln \left( \frac{\sqrt{1-x} + \sqrt{1-x-\lambda}}{ \sqrt{1-x} - \sqrt{1-x-\lambda} } \right) = \\
& \frac{q_q^2 \lambda}{16 \pi f_a^2}&\left(-\sqrt{4-2 \lambda }+
(\lambda -2) \log \left(\sqrt{2}-\sqrt{2-\lambda }\right)
 +2 \log
   \left(\sqrt{2-\lambda }+\sqrt{2}\right)-\frac{1}{2} \lambda  \log (\lambda ) \right). \nn
\eea
Finally, we  need to integrate $\langle \sigma v \rangle$ over the  photon spectrum
using the thermal number density, 
$$
dn = \frac{g}{2 \pi^2} \frac{E_{\gamma}^2 dE_{\gamma}}{ e^{\frac{E_{\gamma}}{T}} - 1},
$$
with $g=2$. Using $E_{\gamma} = \frac{2 m_q^2}{E_a \lambda}$, we can rewrite
$
\frac{E_{\gamma}^2 dE_{\gamma}}{e^{E_{\gamma}/T} - 1} = \frac{d\lambda}{e^{\frac{2 m_q^2}{E \lambda T}} - 1} \frac{8 m_e^6}{E^3 \lambda^4} \, ,
$
with $\lambda$ running from 0 to 2.
We then end up with,
\bea
\Gamma = \langle n \sigma v \rangle & = & \frac{g}{2 \pi^2} \frac{m_q^6}{E_a^3} \frac{q_q^2}{2 \pi f_a^2} \int_{0}^{2} d \lambda
\frac{1}{e^{\frac{2 m_q^2}{E_a \lambda T}} - 1} \frac{1}{\lambda^3} \left(-\sqrt{4-2 \lambda }+
\right. \nn \\
&+& \left.
2 \log
   \left(\sqrt{2-\lambda }+\sqrt{2}\right)
-(2-\lambda ) \log \left(\sqrt{2}-\sqrt{2-\lambda }\right)
 -\frac{1}{2} \lambda  \log (\lambda ) \right) \label{eq:Gamma}
 \, ,
\eea
which can be evaluated numerically as a function of the relevant parameters.

\subsection{Bounds  on highly relativistic axions from Big Bang nucleosynthesis} \label{sec:bounds}

Equation \eqref{eq:Gamma} gives  the scattering rate of energetic axions off thermal photons, and in this section we will show how this scattering rate --- despite its relative smallness in relation to the Hubble parameter in the early universe --- can be large enough to affect  the primordial $^4$He abundance.

After the decay of the modulus $\Phi$ into Standard Model degrees of freedom and axions, the number density of axions is 
diluted with the expansion of space-time. Occasional scattering of axions  off the thermal plasma into `primary' scattering products such as $q \bar q$ or quark/anti-quark pairs, quickly thermalize  by producing a large number of electromagnetic (i.e.~$\gamma$, $e^{\pm}$) or hadronic (e.g.~$p$, $n$) `secondary' particles. The secondaries  may in turn affect the light element abundances in the early universe through a number of different processes, as described in detail in e.g.~\cite{Ellis:1990nb, Dimopoulos:1988ue, Jedamzik:2006xz, Kawasaki:2004qu}.  This way, relativistic axions may   in principle be constrained by Big Bang Nucleosynthesis bounds on nuclear abundances.

\subsubsection{Inelastic axion scattering and decaying neutral particles}
 Unfortunately, to our knowledge there exists no dedicated study of what constraints may be inferred from light element yields during  BBN on a highly relativistic species which  scatters inelastically off the thermal plasma. However, there are well-known constraints  on massive  particle  species, here denoted  $X$, \emph{decaying} with life-time $\tau_X$ during and after BBN. These bounds are often quoted in terms of
 \be
 \epsilon_X \equiv M_X \xx\, ,
 \ee
 where $\xx = n_X/n_{\gamma}$ is the relative abundance of  $X$-particles to photons prior to the decay, and $M_X$ is the mass of $X$. The quantity $\epsilon_X$ measures the total energy deposited per photon once the entire species $X$ has decayed. Axion scattering off the thermal plasma differs from a decaying particle in four ways:
 \begin{enumerate}
\item The scattering rate, and thereby the effective `life-time' ($\tau = \Gamma^{-1}$), depends on the temperature of the thermal photons as well as the energy of the axion, which both decrease with time. This gives rise to a particular time-dependence for the scattering rate into  each primary final state.
\item While the decay products of a decaying massive particle are traditionally assumed to have negligible  total momentum with respect to the rest frame of the plasma, the primaries of axion scattering will necessarily appear boosted with respect to this frame. Yet,  the average total momentum  over many inelastic axion scatterings will vanish, and we expect this  point to be of less importance to the applicability of the bounds.
\item The role of the factor of  $M_X$ in $\epsilon_X$ which determines the total energy of the decay products is in the case of axion scattering played by the time-dependent axion energy $E_a \gg T$.
\item
Finally, while the number of $X$-particles is greatly reduced at times $t/\tau_X \gg1$, the number density of axions per comoving volume does not change by much for the scattering rates that we will discuss. In other words, axion scatterings are relatively  rare events, and most relativistic axions pass through BBN unperturbed.
\end{enumerate}

Despite these obvious differences, we find that at the level of the energy deposition history --- which is in the end what is constrained by the BBN analysis --- each axion scattering channel can be well approximated by a species of decaying particle. Thus, the effects of the axion during BBN can be modelled by a \emph{collection} of decaying particle species,  each decaying only to a particular final state $q_i \bar q_i$ with  life-time $\tau_i$ and deposited energy (per photon) $\epsilon_i$. The corresponding hadronic branching ratio for each of these scattering modes will be either 1 (for quark and heavy lepton primaries) or $\sim 10^{-3}$ (for electron primaries).

In this section, we impose known bounds on the energy deposition of each individual such species during BBN, and interpret the bound as a constraint on the axion decay constant and modulus mass $m_{\Phi}$. These bounds are necessarily conservative as the effect of each possible primary is considered separately, yet we will see how they can give important constraints on the axion decay constant.

Specifically, constraints on a neutral decaying particle during (and after) BBN are conveniently phrased in terms of bounds on
\be
\epsilon_X = - \int_{0}^{\infty} dt' M_X \dot{\xi}_X(t') =   \int_{0}^{\infty} dt' M_X |\Gamma_X| \xi_X(t') \, ,
\ee
given the life-time $\tau_X$. For axion scattering into a particular final state, here denoted $q \bar q$, the corresponding bounds can be phrased in terms of
\be
\epsilon_a =    \int_{t_0}^{\infty} dt' E_a(t') |\Gamma_{a\gamma \rightarrow q \bar q}(t')| \xi_a(t') \, ,
\ee
where $\xa = n_a/n_{\gamma}$ denotes the relative number density of axions to photons.  Here $t_0$ denotes an initial time before which no bounds can be placed on the energy deposition, and during radiation domination, $E_a(t') = E_0 \left(\frac{t_{rh}}{t' }\right)^{1/2}$. Thus, the deposited energy from axion scattering into $q \bar q$-pairs from the time $t_0$ up to $t$ is given by,
 \be
\epsilon_a(t) = E_0~\int_{t_0}^{t} dt' \left(\frac{t_{rh}}{t' }\right)^{1/2} |\Gamma_{a\gamma \rightarrow q \bar q}(t')| \xi_a(t') \, . \label{eq:epsa}
\ee
Over a period in which the effective number of species remains the same,
the temperature of the photon gas depends on time as $T = T_{rh} \sqrt{t_{rh}/t}$, and the re-heating temperature is determined from the mass of the decaying modulus (and thereby $E_0$) as in equation \eqref{eq:Trh}.
\begin{figure}
\begin{center}
 \includegraphics[width= .85 \textwidth]{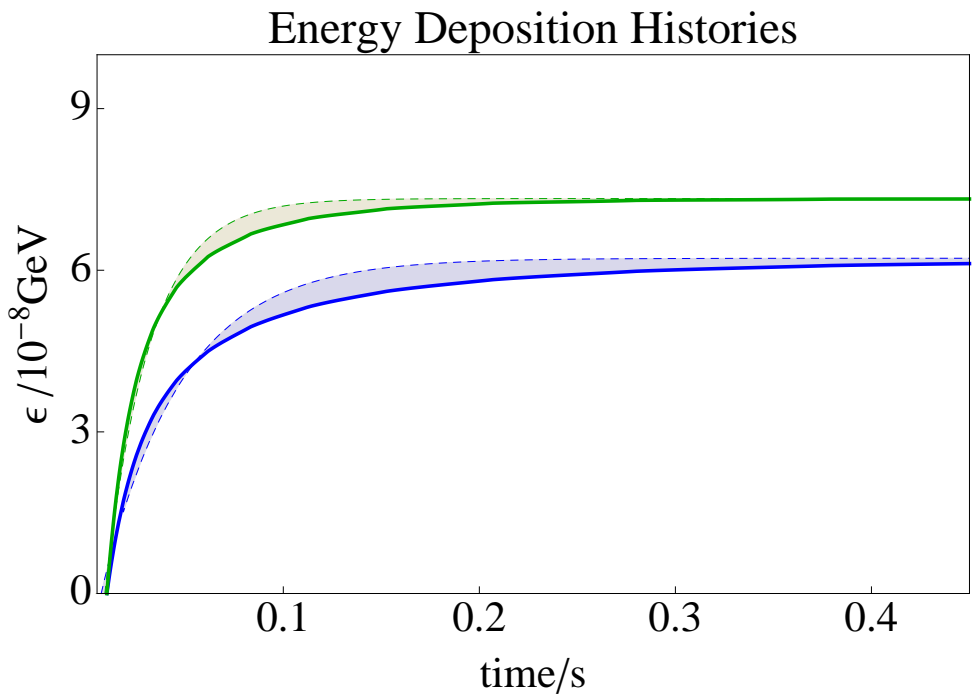}
 \end{center}
 \caption{Energy deposition histories $\epsilon_a(t)$ for axion inelastic scattering to bottom quarks (solid lines) as compared to the best-fit profile of $\epsilon_X(t)$ for a decaying particle species (dashed lines). For both solid curves, $f_a = 10^9$ GeV and $\Delta N_{eff} = 0.57$. In blue, $m_{\Phi}=5\cdot 10^6$ GeV, and in green  $m_{\Phi}=5\cdot 10^7$ GeV.
 }
 \label{fig:Edepos}
 \end{figure}

Thus, the energy deposition for axion scattering exclusively into $q \bar q$ can be found explicitly, and an example thereof is shown in figure \ref{fig:Edepos} for scattering into bottom quarks. From this figure, we note that the energy deposition history from axion scattering is very similar to that of a decaying particle species, $\epsilon_X(t) =  M_X( \xi_X(t_0) - \xx(t))$.  This allows us to map each axion scattering mode into a species of neutral decaying particles with life-time, mass and initial number density determined by optimising the fit of the energy deposition profile to that of the corresponding axion scattering.

However, before turning to the detailed constraints obtained by carefully evaluating the analysis presented in this section, we will now give an order of magnitude estimate of the energy deposition $\epsilon_a$ for the particular case in which $m_{\Phi} = 2\cdot10^{6}$ GeV, and the axions give rise to an excess radiation of $\Delta N_{eff} = 0.57$. In this case, the effective `life-time' for axion-photon scattering into a $b \bar b$ pair is $\tau \approx 0.2$ s, at which time the temperature of the plasma is $T \approx 4$ MeV and $E_a \approx 16$ TeV. For this rough estimate, we evaluate the scattering cross-section by taking $f_a = 10^9$ GeV and $E_{\gamma} = T$ in equation \eqref{eq:sigma}, and only consider axion-photon scattering at a $90$\textdegree angle.  Upon substituting the fermion mass in equation \eqref{eq:sigma} by the bottom quark mass, we find that 
\be
\sigma_{\rm approx} \approx 6 \cdot 10^{-28}~{\rm MeV}^{-2} \, . 
\ee
For this order of magnitude estimate, we note that $\rho_a = \Delta N_{eff} \rho_{\nu}=  \Delta N_{eff} \frac{7}{8} \frac{\pi^2}{15} T^4  = 0.58 (4~{\rm MeV})^4$, where $\rho_{\nu}$ denotes the energy density of a single neutrino species at $T= 4$ MeV. Similarly, we approximate the Hubble parameter to be  $H \approx 7.2 \cdot 10^{-21}$ MeV at this time. The energy deposition from axion-photon scattering into $b \bar b$ quarks may thus be estimated as
\be
\epsilon_a \approx \frac{1}{H n_{\gamma}} \sigma_{\rm approx} {\rho}_a n_{\gamma} \approx 1 \cdot 10^{-8}~{\rm GeV} \, .
\ee
From figure 6 of \cite{Jedamzik:2006xz}, we find that the bound on $\epsilon_X$ for a decaying neutral particle of mass $M_X= 1$ TeV at $t=0.2$ s is given by $\epsilon_X < 4 \cdot 10^{-8}~{\rm GeV}$. Thus, based on this heuristic estimate, we conclude that inelastic axion-photon scattering may deposit significant amounts of energy in the thermal plasma during BBN, and we will now proceed by a more careful evaluation of the corresponding bounds on the axion decay constant.

 \subsubsection{Observational constraints from BBN}
During BBN ($10^{-2} \lesssim t \lesssim 10^2$ seconds), the strongest constrains apply to processes which increase the relative number of neutrons to protons as these neutrons will in turn increase the Helium abundance, $Y_p$.  At the very beginning of this period, $t\lesssim 3 \cdot 10^{-2}$ seconds, even these bounds are ineffective as thermal weak interactions quickly restore the neutron to proton ratio. However, particles decaying later during BBN may be constrained from bounds on overproduction of $^4$He.

The primordial Helium abundance can be  inferred both from spectroscopy of  extra galactic H$_{\rm II}$ regions by extrapolation to zero metallicity, as well as from fits of cosmological models of  the temperature anisotropies of the CMB and the large-scale structure of the universe (for a recent review, see \cite{Steigman:2010pa}).
Recent spectroscopic studies have found $Y^{BBN}_p = 0.2565 \pm0.0010 {\rm (stat)} \pm 0.0050 {\rm (syst)}$ \cite{IT:2010},  $Y^{BBN}_p = 0.2528 \pm0.0028$  \cite{AOS:2010},  and $Y_p^{BBN} = 0.2534 \pm 0.0083$ \cite{AOS:2011}. These values are  compatible within $1\sigma$ to those inferred from the recent Planck experiment, which estimated $Y_p^{BBN} = 0.266\pm0.021$ for the $\Lambda$CDM model extended with $Y^{BBN}_p$ as a free parameter, and $Y_p^{BBN} = 0.254^{+0.041}_{-0.033}$ for $\Lambda$CDM extended with both $Y_p^{BBN}$ and $N_{eff}$ (the latter parameter  is then determined to $N_{eff} = 3.33^{+0.59}_{-0.83}$). In this paper, we will use the constraints on decaying neutral particles from  \cite{Jedamzik:2006xz} which are  based on the  conservative bound of $Y_p<0.258$. 



\begin{figure}
\begin{center}
\includegraphics[width=.90 \textwidth]{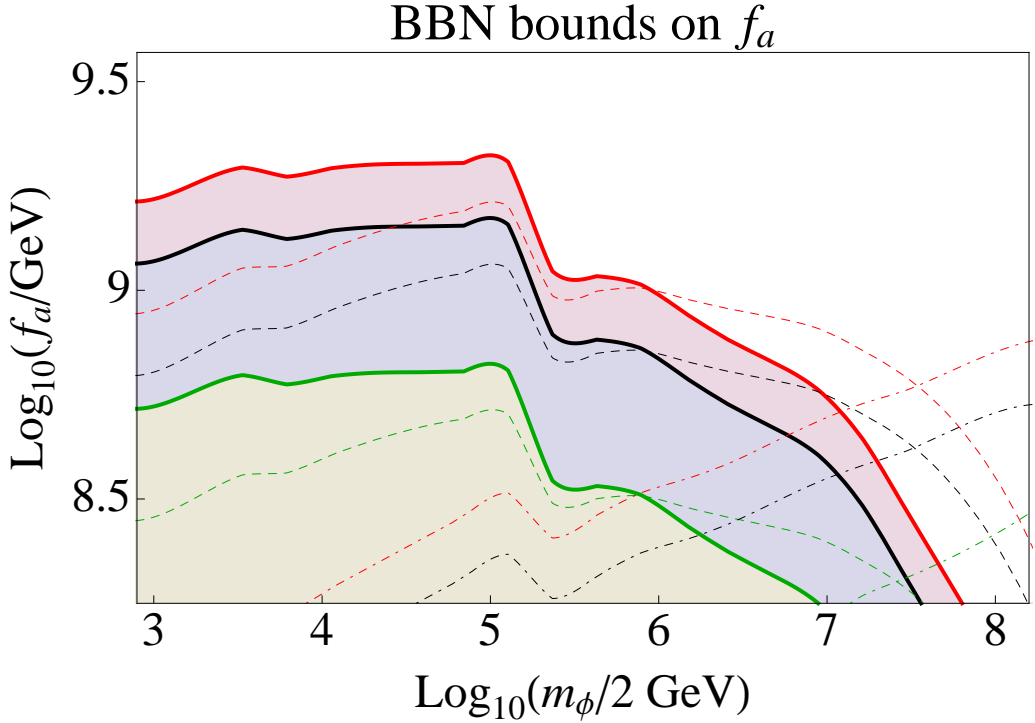}
 \end{center}
 \caption{Conservative bounds on the axion decay constant as a function of the initial axion energy obtained by considering the effects of the decay channels to $b\bar b$ (solid), $c \bar c$ (dashed), and $s \bar s$ (dot-dashed) separately. The  red, black and green curves correspond to
 $\Delta N_{eff} = 0.1, 0.5$ and $1$, respectively, and the areas below the curves are excluded from the constraints from \cite{Jedamzik:2006xz} due to overproduction of $^4$He during BBN. 
 }
 \label{fig:bounds}
 \end{figure}

Given the resulting constraint on $\xi_X$ for a given life-time, a (conservative) bound on the energy deposited by the scattering axion into each decay channel may be derived. The resulting constraints as obtained separately for  the $b \bar b$, $c \bar c$ and $s \bar s$ scattering channels are shown in figure \ref{fig:bounds}. We note that these constraints probe an interesting region of the axion parameter space, and that dedicated studies of inelastic axion scattering during BBN may well be able to derive even stronger bounds on $f_a$.

\section{Dark Matter Axiogenesis} \label{sec:dm}

One popular model of dark matter is that it consists of the lightest supersymmetric particle (LSP) whose stability is
protected by R-parity. Conventionally, the abundance of the LSP is determined via a thermal freeze out from the
ambient plasma at a temperature of approximately $T_{freeze-out} \simeq m_{LSP}/20$.

\begin{figure}
\begin{center}
\includegraphics[width=8cm]{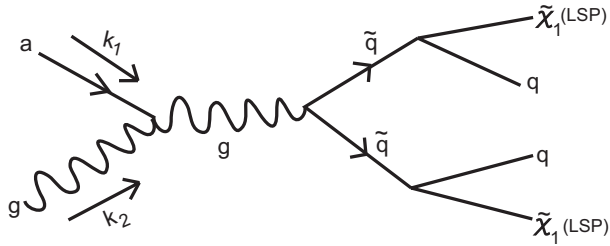}
\end{center}
\vspace{-5ex}
\caption{Dark matter production from axion scattering. Here the axion scatters off a gluon, leading to
pair production of heavy squarks which then cascade decay to produce the LSP.}
\label{axiogenesis}
\end{figure}

However, the scenario considered here offers a radically new production mechanism.
Even when the thermal plasma has $T < T_{freeze-out} \ll m_{LSP}$,
energetic axionic dark radiation can directly produce supersymmetric particles through scattering off the ambient plasma.
The produced susy particles will undergo a rapid cascade decay to the LSP,
which will remain as a stable dark matter relic. A sample process leading to dark matter production
is shown in figure \ref{axiogenesis}.
The rate of dark matter production is set by the axion scattering rate. This is an entirely non-thermal
process, and can give significant LSP production at temperatures $T  \ll T_{freeze-out}$.

Let us first give the basic scaling estimates for the dark matter abundance produced (a proper calculation follows).
For the estimate we assume the axion-plasma center of mass
energy is above threshold for producing susy particles and neglect mass effects. We also use only the `universal' $a \gamma \gamma$ or $a g g$
vertex shown in figure \ref{axiogenesis}, so there is no mass suppression of the axion-matter coupling.

The density of plasma particles is $n \sim T^3$ and the axion scattering cross-section is
 $ \sigma \sim \left( \frac{\alpha}{4 \pi} \frac{1}{f_a} \right)^2$.
The total scattering rate of an axion with the ambient plasma is then
\be
\Gamma = \langle n \sigma v \rangle \sim \left( \frac{\alpha}{4 \pi} \right)^2 \frac{T^3}{f_a^2} \, .
\ee
As during radiation domination $T \sim t^{-1/2}$, the scattering rate
$\Gamma \sim t^{-3/2}$ and it is clear that the total number of scattering events $\mc{N}_{total} = \int \Gamma dt$ is dominated by
those occurring at early times. As the Hubble scale is
$H \sim \frac{T^2}{M_{Pl}}$, the fraction of axions that interact in a Hubble time is
$$
\frac{\Gamma}{H} \sim \frac{T M_{Pl}}{f_a^2} \left( \frac{\alpha}{4 \pi} \right)^2. \, 
$$
As this is dominated by early times, in the instantaneous-reheating approximation,
we simply evaluate this at $T \sim T_{reheat}$ to get an estimate of the total fraction of axions that scatter.
Each decaying modulus of mass $m_{\Phi}$ generates either two axions of energies $E_a = m_{\Phi}/2$ or
$\sim m_{\Phi}/T_{reheat}$ thermalised Standard Model particles. If the axion branching ratio is $B_a$, the
axion abundance is
$$
Y_{axion} \equiv \frac{n_{axion}}{s} \sim \frac{B_a T_{reheat}}{m_{\Phi}}.
$$
As each axion scattering event to R-parity odd particles produces two dark matter particles, the resulting dark matter abundance is
$$
Y_{dm} \sim Y_{axion} \frac{2 \Gamma}{H} \sim B_a \frac{T_{reheat}^2 M_{Pl}}{m_{\Phi} f_a^2} \left( \frac{\alpha}{4 \pi} \right)^2.
$$
While this neglects factors of $g_{*}, \pi$ etc, it is clear that for reasonable parameter values
this can be a phenomenologically interesting source of dark matter (note that $Y_{dm} \sim 1.5 \ti 10^{-12}$ for a 300 GeV LSP).

\begin{figure}
\begin{center}
\includegraphics[width=14cm]{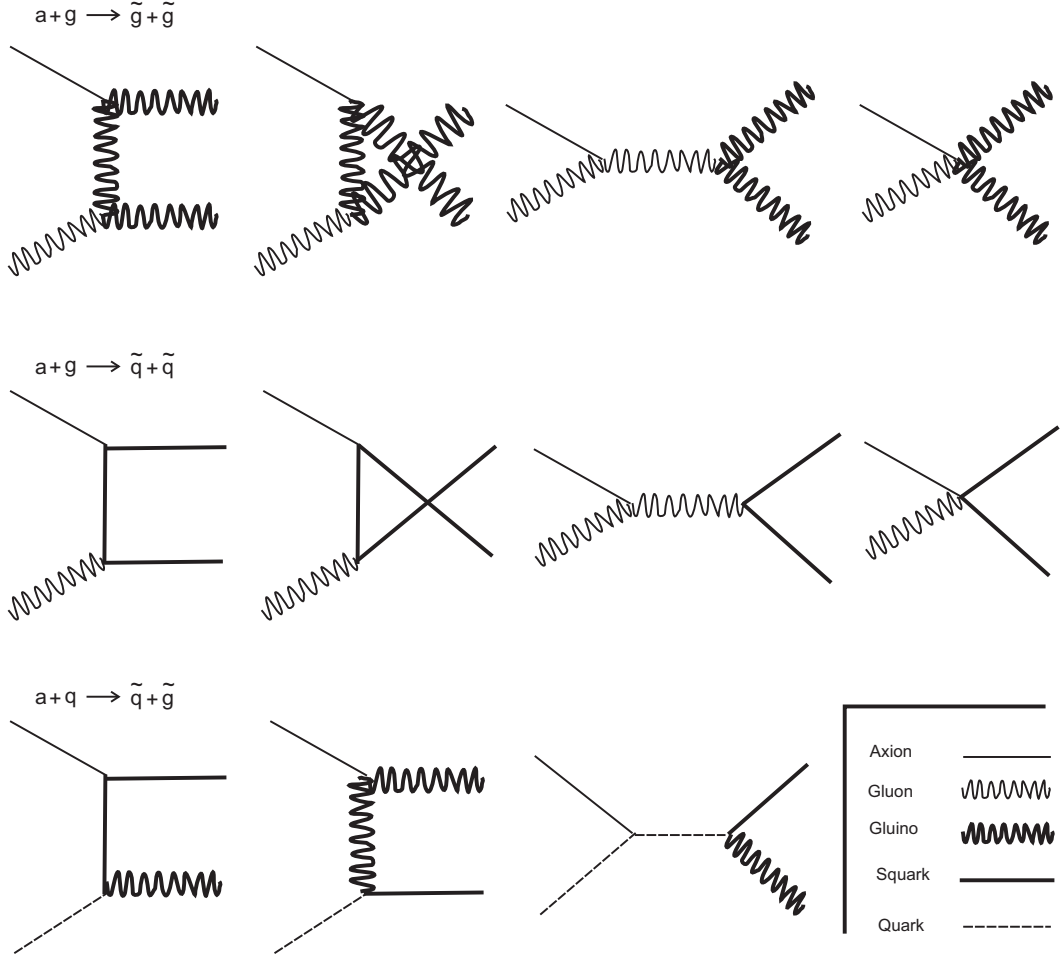}
\end{center}
\vspace{-5ex}
\caption{Feynman diagrams that lead to LSP production from axion scattering off the ambient plasma. We have shown diagrams that involve strong
interactions but there are also similar diagrams involving weak or electromagnetic scattering off photons or leptons.}
\label{AllDiagrams}
\end{figure}

Overall, any source of axion-plasma interaction that produces susy particles can produce dark matter. Such diagrams are shown
in figure \ref{AllDiagrams}.
We will return to the full set of diagrams of figure \ref{AllDiagrams} in future work.
Here however we focus only on the universal diagrams involving s-channel photons or gluons. These are universal
as the coupling $a g g$ is necessarily present for a QCD axion, whereas any
interactions with quarks or squarks are more model dependent
with unspecified coefficients.

\subsection{Scattering via Photons}
\label{PhotonScattering}

We consider the scattering process $a + \gamma \to \tilde{q} \tilde{q}^{*}$ via an s-channel photon.
We treat the axion as massless, and give the squarks a mass of $m_{\tilde{q}}$.
In section \ref{GluonScattering}
we will consider the analogous process $a + g \to \tilde{q} \tilde{q}^{*}$. The $a \gamma \gamma$ interaction is
\be
\label{atwophoton}
\mc{L}_{a\gamma \gamma} = a g_{a\gamma \gamma} {\bf{E} \cdot \bf {B}} \equiv \frac{1}{8} a g_{a \gamma \gamma} \epsilon_{\mu \nu \lambda \rho} F^{\mu \nu} F^{\lambda \rho}.
\ee
The resulting amplitude is,
\be
\mc{M} = e g_{a \gamma \gamma} \epsilon_{\mu \nu \lambda \rho} \frac{k_2^{\mu} k_1^{\lambda} \epsilon^{\nu} (k_3 - k_4)^{\rho}}{(k_1 + k_2)^2},
\ee
where $k_1$ and $k_2$ are the incident 4-momenta of the axion and photon, and $k_3$ and $k_4$
the 4-momenta of the outgoing squarks or sleptons. In the center of mass frame we can without loss of generality write
\bea
k_1 & = & (E_{CM}, 0, 0, E_{CM}) \, ,\nn \\
k_2 & = & (E_{CM}, 0, 0, -E_{CM})\, , \nn \\
k_3 & = & (E_{CM},0, p \sin \theta, p \cos \theta)\, , \nn \\
k_4 & = & (E_{CM},0, -p \sin \theta, -p \cos \theta)\, , \nn \\
\epsilon & = & (0, \cos \phi, \sin \phi, 0) \, , \nn
\eea
giving
\be
\vert \mc{M} \vert^2 = e^2 g_{a \gamma \gamma}^2 p^2 \cos^2 \phi \sin^2 \theta \, .
\ee
Averaging over the initial photon polarisation then gives
\be
\vert \mc{M} \vert^2 = \frac{e^2 g_{a \gamma \gamma}^2}{2} p^2 \sin^2 \theta \, .
\ee
If the lab frame 4-momenta are
\bea
k_a & = & (E,0,0,E) \, , \nn \\
k_{\gamma} &  = & (E_{\gamma}, 0, E_{\gamma} \sin \theta, E_{\gamma} \cos \theta) \, , \nn
\eea
then the resulting cross-section, integrated over final state phase space, is
\be
\sigma v = \frac{1}{4 E E_{\gamma}} \frac{e^2 g_{a \gamma \gamma}^2}{24 \pi} \sqrt{1 - \frac{ 2m^2}{E E_{\gamma} (1-x)}}
\left( \frac{E E_{\gamma}}{2} (1-x) - m^2 \right) \, ,
\ee
where $x = \cos \theta$. We finally average over the relative angle between the photon and axion,
$
\langle \sigma v \rangle = \half \int_{-1}^{1-\lambda} \sigma v dx.
$
In terms of $\lambda = \frac{ 2m^2}{E E_{\gamma}}$, this is
\be
\langle \sigma v \rangle = \frac{1}{128} \frac{e^2 g_{a\gamma \gamma}^2}{24 \pi} \left( 2 (4 - 5\lambda) \sqrt{4 - 2 \lambda}
+ 6 \lambda^2 \ln \left( \sqrt{2} + \sqrt{2 - \lambda} \right) - 3 \lambda^2 \ln \lambda \right) \, ,
\ee
for the x-section $\langle \sigma v \rangle$ of an axion of energy $E$ incident on a single photon of energy $E_{\gamma}$.
We finally sum over the spectrum of thermal photons by integrating
$$
\frac{g^2}{2 \pi^2} \int \frac{E_{\gamma}^2}{e^{E_{\gamma}/T} - 1} d E_{\gamma} = \frac{g^2}{2 \pi^2} \int_0^2 d \lambda
\frac{1}{e^{\frac{2 m^2}{E \lambda T}} - 1} \frac{8 m^6}{E^3 \lambda^4} \, .
$$
The scattering rate to a pair of charged scalars is finally
\bea
\label{abcd}
\Gamma = \langle n \sigma v \rangle & = &  \frac{g^2}{2 \pi^2} \int_0^2 d \lambda
\frac{1}{e^{\frac{2 m^2}{E \lambda T}} - 1} \frac{8 m^6}{E^3 \lambda^4} \ti \frac{1}{128} \frac{e^2 g_{a\gamma \gamma}^2}{24 \pi} \left( 2 (4 - 5\lambda) \sqrt{4 - 2 \lambda} \right. \nn \\
& & \left. + 6 \lambda^2 \ln \left( \sqrt{2} + \sqrt{2 - \lambda} \right) - 3 \lambda^2 \ln \lambda \right) \, .
\eea
We obtain the total scattering rate to sleptons and squarks by summing over all species. For simplicity, we assume a common
mass scale for all squarks and sleptons. Writing $e = q \sqrt{\frac{4 \pi}{137}}$, we have
\be
\Gamma_{total} = \underbrace{6 \ti \Gamma(q=1)}_{\hbox{sleptons}} + \underbrace{6 \ti 3 \ti \Gamma(q=2/3)}_{\hbox{up squarks}} +
\underbrace{6 \ti 3 \ti \Gamma(q=1/3)}_{\hbox{down squarks}} \, .
\ee
As each axion scattering event produces pairs of susy particles, each of which undergoes a rapid cascade decay down to the LSP,
the overall LSP production rate is
\be
\Gamma_{LSP} = 2 \ti \Gamma_{total} \, .
\ee

\subsection{Scattering via Gluons}
\label{GluonScattering}

There is also a kinematically identical diagram where the s-channel photon is replaced by a gluon.
This differs only by the colour factors and the conventional notation for axion-photon or axion-gluon couplings.
The $a g g$ vertex comes from the interaction
\be
\label{atwogluon}
\frac{a}{f_{PQ}} \frac{\alpha_s}{16 \pi} \epsilon_{\alpha \beta \gamma \delta} G^{\alpha \beta}_a G^{\gamma \delta}_a \in \mc{L} \, .
\ee
By comparison of eq. (\ref{atwogluon}) with eq. (\ref{atwophoton}), we see that we have to replace
$$
(g_{a \gamma \gamma})_{photon} \hbox{ with } \left( \frac{1}{f_{PQ}} \frac{\alpha_s}{2 \pi} \right)_{strong} \,
$$
and
$$
-ie (k_3 - k_4)^{\mu} \hbox{ with } -ig_s (k_3 - k_4)^{\mu} t^b,
$$
where $g_s$ is the strong coupling and $\alpha_s = \frac{g_s^2}{4 \pi}$.
Summing over the final squark colour states, in eq. (\ref{abcd}) we replace
$$
e^2 g_{a \gamma \gamma}^2 \hbox{ with } \left( \frac{1}{f_{PQ}} \frac{\alpha_s }{2 \pi} \right)^2 \frac{g_s^2}{2} \, .
$$
As all possible scalar final states are squarks, we have
\be
\label{GammaLSP}
\Gamma_{total} = 12 \ti \Gamma, {\hbox{ and again }} \Gamma_{LSP} = 2 \ti \Gamma_{total} \, .
\ee

\subsection{Dark Matter Abundances}

In the instantaneous reheating approximation, we will take $\frac{\Gamma(T_{reheat})}{H(T_{reheat})}$ as a proxy for the total fraction of axions that scatter. The resulting dark matter
number density is
\be
n_{LSP} = 2 \ti n_{axion}(T_{reheat}) \ti \frac{\Gamma(T_{reheat})}{H(T_{reheat})} \, .
\ee
The axion number density is given by eq.~\eqref{eq:n0} which we repeat here,
\be
n_{axion,0} = B_a \ti \frac{3 H(T_{reheat})^2 M_{Pl}^2 }{E_a} \, ,
\ee
with $B_a$ the modulus branching ratio to axions.

As the entropy density is $s = \frac{2 \pi^2}{45} g_{*} T^3$, the dark matter abundance is given by
\be
Y_{LSP} \equiv \frac{n_{LSP}}{s} = \frac{45}{g_{*} \pi^2} \ti \frac{ n_{axion}(T_{reheat})}{T_{reheat}^3} \ti \frac{\Gamma(T_{reheat})}{H(T_{reheat})} \, .
\ee
$\Gamma(T_{reheat})$ is set by eq. (\ref{GammaLSP}). $H(T_{reheat})$ is determined by taking
\be
\Gamma_{\Phi \to {\hbox{all}}} = \frac{1}{4 \pi} \frac{m_{\Phi}^3}{(M_{Pl}/\kappa)^2} \, ,
\ee
where $\kappa$ is a parameter we scan over.

\begin{figure}
\twographs{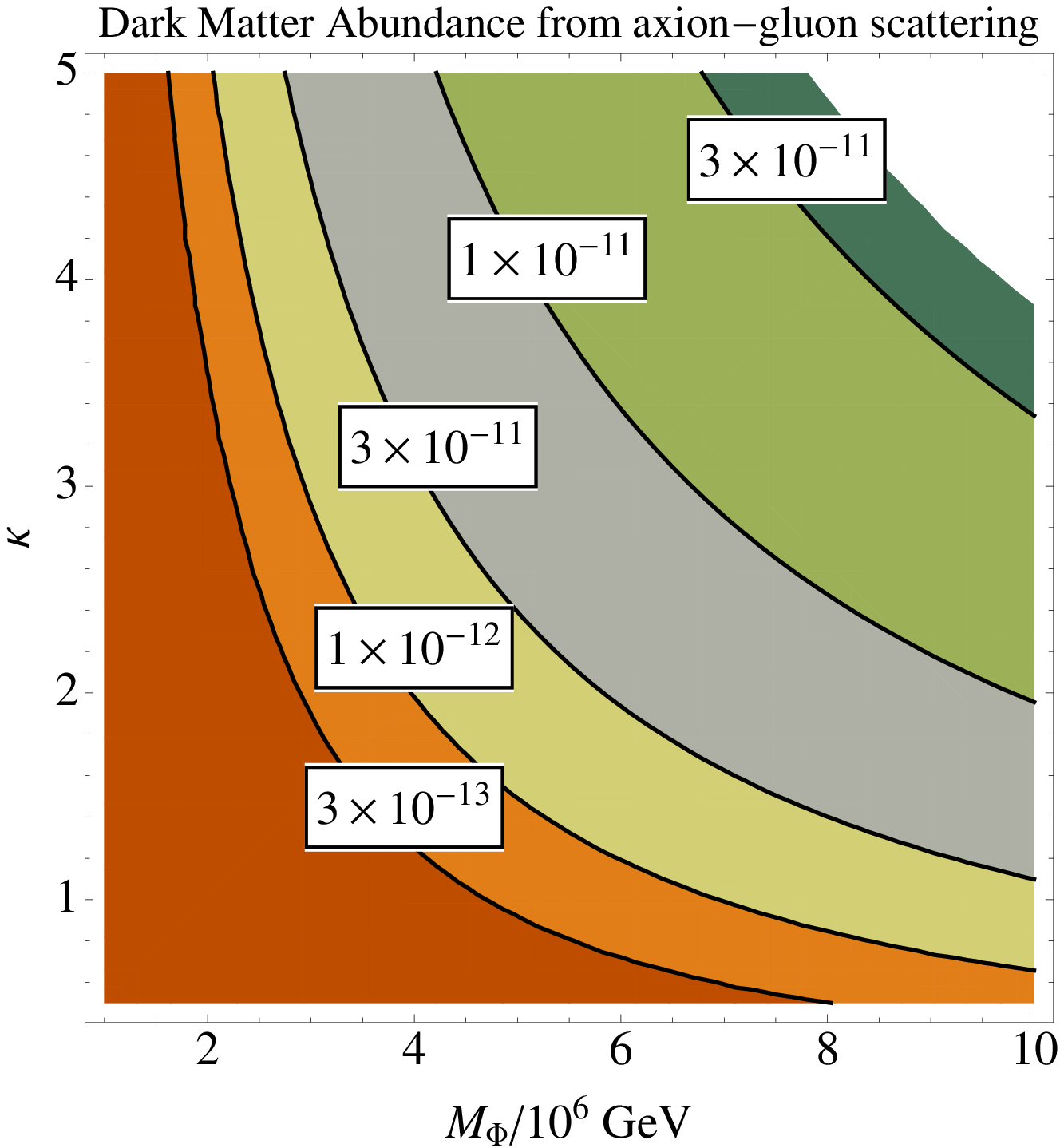}{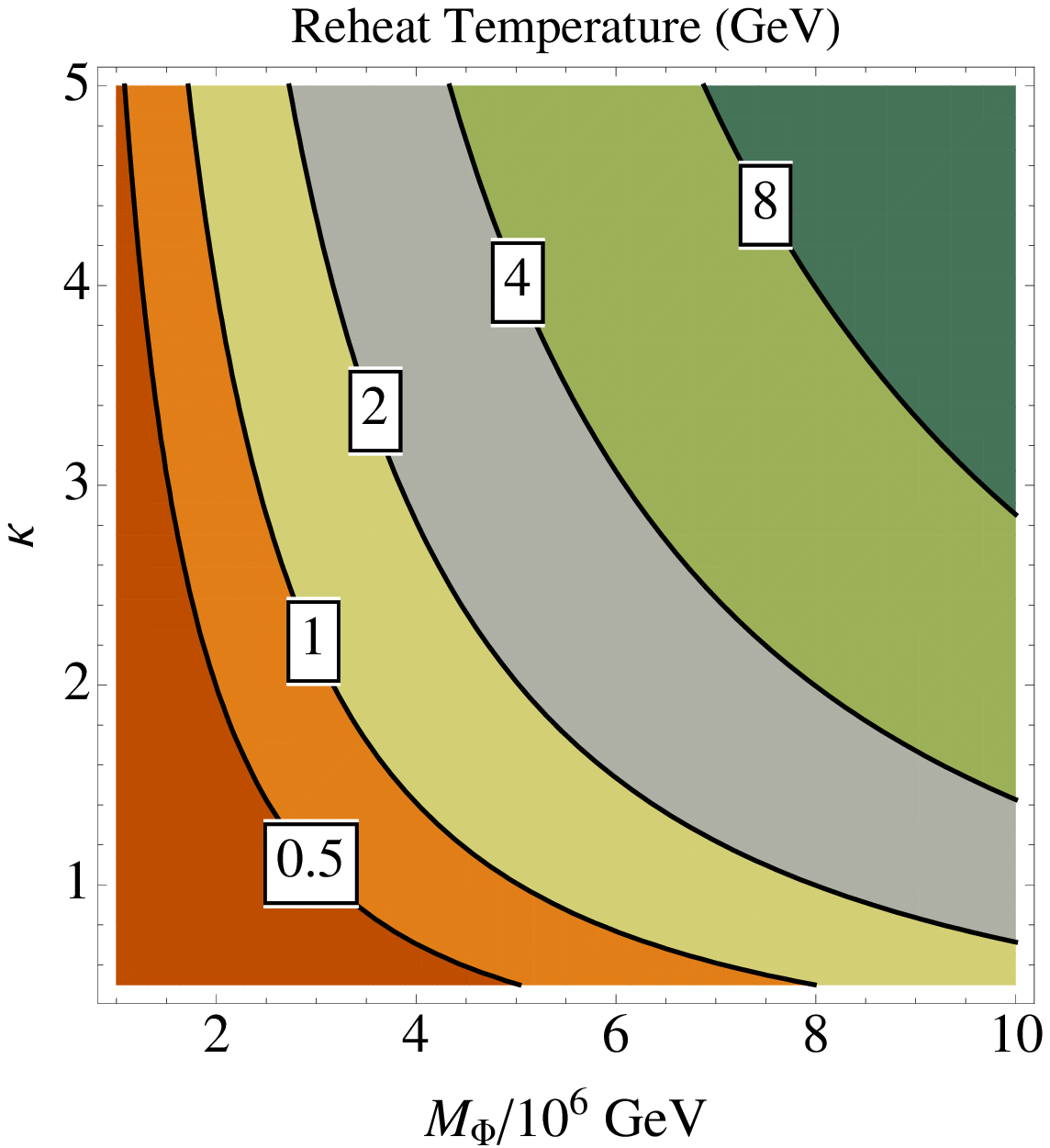}
\caption{The dark matter abundance and reheating temperature as a function of $\kappa$ and $m_{\Phi}$. We have assumed $N_{eff} = 3.62$, $f_{PQ} = 10^9 ~\hbox{GeV}$ and a common squark and slepton mass scale of $1~ \hbox{TeV}$.\label{DMTempPlot}}
\end{figure}

We scan over values of $m_{\Phi}$ between $10^6~ \hbox{GeV}$ and $10^7~ \hbox{GeV}$, and values of $\kappa$ from $0.5$ to $5$.
The results are shown in figures \ref{DMTempPlot}. The figures use $N_{eff} = 3.62$, $f_{PQ} = 10^9~ \hbox{GeV}$ and a common scalar mass
of $1 ~\hbox{TeV}$. We also set $\alpha_s = 0.11$ and use $g_{*} = 61.75$.
It is clear that it is easy to obtain dark matter abundances compatible with observations.

\section{Present Day Cosmic Axion Background Flux} \label{sec:today}
As we have discussed in the previous sections, dark radiation in the form of relativistic axions rarely interacts with its surrounding and the vast majority of the particles  pass through the cosmos unperturbed, leaving a present day population of relic axions which is only diluted by the expansion of space.  In this section, we compute the flux of axionic dark radiation in the current era (hence, `today'), and find that it may even  dominate over the flux of solar axions which are currently searched for by e.g.~the CAST experiment at CERN.  We furthermore point out that the differential axion spectrum and the conversion probability  differs between solar axions and axions arising as dark radiation from moduli decay. 

We re-emphasise here that while we focus for definiteness on axions, the same
spectrum and flux will also hold for any other dark radiation candidate --- for example hidden photons --- that is produced by modulus decay.

Let us first discuss the flux of primordial axions arising from the instantaneous decay approximation for the modulus $\Phi$, thus constituting a mono-energetic population at all times. Immediately after the decay of $\Phi$, the population of axions is given by equation \eqref{eq:n0},
\be
n_{axion, 0} = \frac{B_a \kappa^4}{6 \pi^2 } \frac{m_{\Phi}^5}{M_{Pl}^2} = 3.7\cdot 10^{32}\cdot B_a \kappa^4 \left( \frac{m_{\Phi}}{10^6~{\rm GeV}}\right)^5~{\rm cm}^{-3} \, , \nn
\ee
which we repeat here for clarity. 
Using the expression for $T_{rh}$ from  equation \eqref{eq:Trh}, the entropy at time of reheating is easily found to be,
\be
s = \frac{2 \pi^2}{45} g_*  T_{reheat}^3 = \frac{2 \kappa^3}{45 \pi} g_*^{1/4}
\left( \frac{5}{2}(1-B_a) \right)^{3/4} \frac{m_{\Phi}^{9/2}}{M_{Pl}^{3/2}} \, ,
\ee
where $g_{\star} =g_*(T_{reheat})$.
The axion abundance --- both initially and at the present day --- is then
\bea
Y_{axion} &\equiv& \frac{n_a}{s} = B_a \frac{45}{12 \pi} \frac{\kappa}{g_*^{1/4}}
\left( \frac{2}{5(1-B_a)} \right)^{3/4} \frac{m_{\Phi}^{1/2}}{M_{Pl}^{1/2}} = \\
&=& 3.85 \cdot 10^{-7} \frac{\kappa}{g^{1/4}_{\star}}\frac{B_a }{(1-B_a)^{3/4}} \left(\frac{m_{\Phi}}{10^6~{\rm GeV}} \right)^{1/2} \, .
\eea
With the present day CMB photon density of $n_{\gamma} = 413$ photons per cm$^{-3}$, the axion number density today is given by,
\be
n_a = Y_{axion} s \simeq 7.04 n_{\gamma} Y_{axion}  = 1.12 \cdot 10^{-3}\frac{\kappa}{g^{1/4}_{\star}}\frac{B_a }{(1-B_a)^{3/4}} \left(\frac{m_{\Phi}}{10^6~{\rm GeV}} \right)^{1/2}~{\rm cm^{-3}} \, .
\ee
The corresponding flux of dark radiation axions is then simply given by,
\be
\Phi_a = \frac{n_a c}{4} = 8.39\cdot10^6\frac{\kappa}{g^{1/4}_{\star}}\frac{B_a }{(1-B_a)^{3/4}} \left(\frac{m_{\Phi}}{10^6~{\rm GeV}} \right)^{1/2}~{\rm s^{-1} cm^{-2}} \, .
\label{eq:flux}
\ee
For the example we have considered throughout this paper with $m_{\Phi}= 5 \cdot 10^6$ GeV, $\kappa=1$, $g_{\star}=61.75$ and $B_a =0.145$, the present day flux is,
\be
\Phi_a = 1.09\cdot10^6~{\rm s^{-1} cm^{-2}} \, .
\ee
 This flux may be compared to that of axions created from Primakoff scattering in the sun, which --- contrary to the isotropic dark radiation flux of equation \eqref{eq:flux} --- is always directional and suppressed by two powers of the axion decay constant \cite{RaffeltBook},
 \bea
 \Phi_{solar} & = & \left( g_{a \gamma \gamma} \ti 10^{10} \hbox{GeV} \right)^2 3.54 \ti 10^{11} \hbox{cm}^{-2} \hbox{s}^{-1} \nn \\
& \equiv & 1.91 \ti 10^6 \left( \frac{10^{10} \hbox{GeV}}{f_a} \right)^2 \hbox{cm}^{-2} \hbox{s}^{-1} \, .
 \eea
 The present day flux of dark radiation axions can easily dominate the solar axion flux for
 non-excluded values of the axion decay constant.

Moreover, the spectrum of solar axions differ significantly from that of axions produced from moduli decays. The
differential solar flux spectrum is given by
\be
\frac{d\Phi}{dE} = \left( \frac{\alpha_{em}}{\pi} \right)^2 \frac{1}{f_a^2} \, 4.02 \ti 10^{10} \hbox{cm}^{-2} \hbox{s}^{-1}
\hbox{keV}^{-1} \frac{ (E/ \hbox{keV})^3}{e^{E/(1.08 \, \textrm{\footnotesize{keV}})} - 1} \, ,
\ee
which peaks around $4 \hbox{keV}$.

In the instantaneous decay approximation --- which we will soon go beyond ---
 the ratio of the axion energy to the CMB temperature is easily obtained. Initially, $E_a$ = $m_{\Phi}/2$ and $T_{\gamma} = T_{reheat}$.
The axion energy redshifts with the expansion of the universe as $E_a \sim \frac{1}{R}$, while the photon energies redshift as
$T_{\gamma} \sim \frac{1}{g_*^{1/3} R}$. There are two boosts to the photon energy: from the reheat temperature to the time of neutrino
decoupling, and from neutrino decoupling to now. These give
\be
\frac{T_{\gamma}}{E_a} = \left( \frac{g_*(T_{reheat})}{g_*(T_{\nu \, decoupling})} \right)^{1/3} \left( \frac{11}{4} \right)^{1/3}
\left( \frac{T_{\gamma}}{E_a} \right)_{reheat},
\ee
which evaluates to
\be
\label{abcd}
\frac{T_{\gamma}}{E_a} = \left( \frac{11}{4} \right)^{1/3} \frac{2 \kappa}{\pi} \left( \frac{1}{10.75} \right)^{1/3}
\left( \frac{5(1-B_a)}{2} \right)^{1/4} g_*(T_{reheat})^{1/12} \left( \frac{m_{\Phi}}{M_{Pl}} \right)^{1/2} \, .
\ee

Thus, in the instantaneous decay approximation, the energy of the dark radiation axions today is given by,
\be
E_a(T_{\gamma}) = 1.97  \left( \frac{M_{Pl}}{m_{\Phi}} \right)^{1/2} \frac{T_{\gamma}}{\kappa g_{\star}^{1/12} (1-B_a)^{1/4}} \, ,
\ee
where we have again abbreviated $g_{\star} = g_{\star}(T_{reheat})$. In the example considered above with $m_{\Phi}= 5\cdot 10^6$ GeV, the present day energy of the axions is,
\be
E_a({\rm today}) = 238~{\rm eV} \, ,
\ee
which is an order of magnitude lower than the energy of solar axions.

In the above discussion, we have used the instantaneous reheating approximation, where all moduli decayed at a time
$\tau$. This produces a monoenergetic axion spectrum, with a single initial energy of $E_a = m_{\Phi}/2$ that is subsequently redshifted.
In reality the moduli decay gradually, and the axion energies are redshifted by different amounts depending on whether they come from
early-decaying or late-decaying moduli.

We now want to determine the actual form of this spectrum.
We note that this structure of the axion energy spectrum is essentially fixed by a time $t \sim (\hbox{a few}) \tau$. At this point all the moduli will have decayed and the universe will have transitioned to radiation domination. As the axions are to a good approximation non-interacting,
the energy spectrum now can be obtained simply by redshifting this spectrum.

To determine the structure of the \emph{energy} spectrum we can calculate within a comoving volume: there is then no
dilution in number density due to expansion of the universe. To convert to physical number densities, we simply multiply by a factor
of $R^{-3}$ - which clearly does not affect the form of the energy spectrum.

For a modulus with lifetime $\tau$, the fractional decay rate is
$$
dN = - \frac{N dt}{\tau}, \hbox{ giving } N = N_0 e^{-\frac{t}{\tau}}.
$$
The initial axion energy from a decay at time $t_d$ is always $E_0 = \frac{M_{\Phi}}{2}$, and at time $t>t_d$ is
$$
E_t = E_0 \left( \frac{a(t_{d})}{a(t)} \right).
$$
Axions generated between times $t_d$ and $t_d + dt_d$ then lie in energies between $E_t$ and $E_t + dE_t$, where
$$
dE_t =  \frac{\dot{a}(t_d)}{a(t_d)} E_t dt_d \equiv E_t H(t_d) dt_d,
$$
and the number of axions generated between times $t_d$ and $t_d + dt_d$ is
\be
dN_a = \frac{2 B_a}{\tau} N_0 e^{-\frac{t_d}{\tau}} dt_d =  \frac{2 B_a}{\tau} N_0 e^{-\frac{t_d}{\tau}} \frac{1}{E_t H(t_d)} dE_t \, .
\ee
At any time $t$, we then have the axion number spectrum
\be
\label{AxionSpectrum}
\frac{dN}{dE_t} =  \frac{2 B_a}{\tau} N_0 e^{-\frac{t_d}{\tau}} \frac{1}{E_t H(t_d)},
\ee
where $t_d$ and $H(t_d)$ are implicitly related to $E_t$ by $E_t = E_0 \left( \frac{a(t_{d})}{a(t)} \right)$.

\begin{figure}[]
\begin{center}
\includegraphics[width=.85 \textwidth]{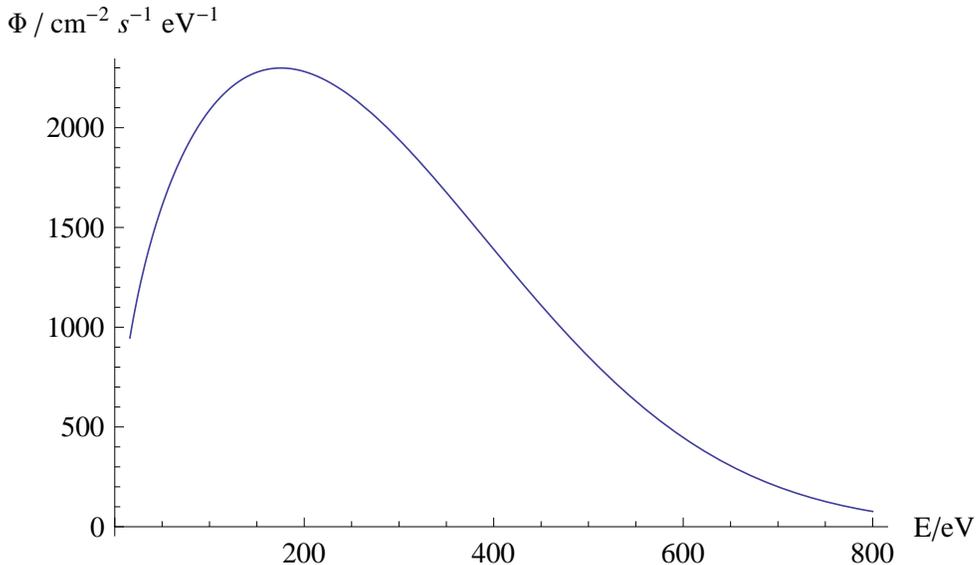}
\end{center}
\vspace{-5ex}
\caption{A typical axion flux per square centimetre per second, with $N_{eff} = 3.62$. The integrated flux is $0.96 \ti 10^6~ \hbox{cm}^{-2} \hbox{s}^{-1}$. The precise location of the energy peak depends on the value of $m_{\Phi}$. The cutoff at low energies is a numerical
artefact from the spectrum computation.
 }
\label{fig:AxionSpectrum}
\end{figure}

The solution of eq. (\ref{AxionSpectrum}) requires the determination of $a(t)$. This can be found numerically as
the modulus gradually decays and the universe transitions from matter domination to radiation domination.
The relevant equations are
\bea
\dot{\rho}_{\Phi} + 3 H \rho_{\Phi} & = & - \frac{\rho_{\Phi}}{\tau} \, , \nn \\
\dot{\rho}_{\gamma} + 4 H \rho_{\gamma} & = & \frac{\rho_{\Phi}}{\tau}\, , \nn \\
H & = &  \sqrt{\frac{\rho_{\Phi} + \rho_{\gamma}}{3 M_{Pl}^2} }\, . \nn
\eea
Using the numerical solution of $a(t)$, the resulting form of the axionic flux is shown in figure \ref{fig:AxionSpectrum}.
The spectrum is normalised to $N_{eff} = 3.62$. The exact location of the energy peak relative to the CMB temperature
depends on $m_{\Phi}$, and a precise determination of this requires an analysis of Standard Model
reheating beyond the instantaneous decay approximation.
However for $m_{\Phi} \sim 10^6 \div 10^7 \hbox{GeV}$ it will always
be more energetic than the CMB peak by an approximate factor of  $10^6$.

We can obtain analytic approximations using the simplifying assumption of either continual radiation domination or continual matter
domination. These assumptions are accurate for either late modulus decays (radiation domination) or early modulus decays
(matter domination).

For radiation domination, as $a \sim t^{1/2}$,
we can relate $t_d$ and $E_t$ by
\be
\frac{t_d}{t} = \left( \frac{E_t}{E_0} \right)^2.
\ee
giving a differential axion number density of
\be
\label{axionspectrum}
\frac{dn(t,E_t)}{dE_t} = \frac{2 B_a N_0}{\tau E_t} e^{- \frac{t}{\tau} \left( \frac{E_t}{E_0} \right)^2} 2 t \left( \frac{E_t}{E_0} \right)^2.
\ee
Continual matter domination with $a \sim t^{2/3}$ gives
\be
\label{axionspectrum2}
\frac{dn(t,E_t)}{dE_t} = \frac{2 B_a N_0}{\tau E_t} e^{- \frac{t}{\tau} \left( \frac{E_t}{E_0} \right)^{3/2}} \frac{3 t}{2}
\left( \frac{E_t}{E_0} \right)^{3/2}.
\ee

This spectrum is characteristic of dark radiation from moduli decays, and may be searched for by experiments like CAST that look for
axion-like particles through conversion in a magnetic field. We note that since the Cosmic Axion Background flux arising from modulus decay is independent of $f_a$, the axion-to-photon generation rate in a magnetic field scales as  $f_a^{-2}$, as compared to the   $f_a^{-4}$ fall-off of the axion-to-photon generation rate for solar axions.

\section{Conclusions}
\label{sec:concl}
In this article, we have studied the rich and interesting cosmophenomenology of axionic dark radiation. We have focused on three disparate areas: additional energy
injection at BBN from axion-matter scattering, the late-time production of dark matter through scattering of axions into susy particles, and
the relic Cosmic Axion Background flux today, which can be comparable to or
larger than the solar axion flux. Our main point
is that there is a rich phenomenology associated to axionic dark radiation, which is by no means sterile.

There are several clear directions for future work, one of which is the extension of our computation in section \ref{sec:dm} to a larger number of processes  for axion scattering into dark matter. In this initial study we principally
considered the simplest case of universal s-channel gluon scattering. There are also many other diagrams, enumerated in figure
\ref{AllDiagrams}, which we did not consider. These diagrams may in fact be more important, as the $a \gamma \gamma$ or $a g g$ vertex is generally suppressed by an extra factor of $\frac{\alpha}{4 \pi}$ compared to axion-matter vertices. Axion-matter-matter couplings do have a suppression by the mass of the relevant matter particle, but this should be less relevant once the centre of mass energy is comparable to the particle masses.

Furthermore, we have emphasised that bounds from overproduction of $^4$He during BBN place interesting constraints on the axion decay constant and the reheating temperature. While our bounds in section \ref{sec:bounds} were derived for each decay channel separately and therefore constitute conservative bounds, a full, inclusive, analysis of the effects of dark radiation during BBN is likely to find even stronger constraints.

In section 5 we pointed out that  dark radiation from
modulus decays has a characteristic isotropic and non-thermal spectrum at an energy scale around six orders of magnitudes higher than
that of the CMB. While we have focused on axions, the same spectrum would apply to other dark radiation candidates such as hidden photons, which
may also interconvert to photons in the presence of a magnetic field.
If it exists, this spectrum represents a Cosmic Axion Background which gives an image of the universe at a time $t \sim 10^{-6} ~\hbox{s}$, and its 
significance is obvious. As the flux is not dissimilar to the solar axion flux
in magnitude, but isotropic in direction, it represents an exciting and worthwhile  target for experiments such as CAST that rely on axion-photon conversion
within a magnetic field.

\section*{Acknowledgments}

JC is funded by a Royal Society University Research Fellowship and by the European Research Council starting grant `Supersymmetry Breaking in String Theory'.
DM was supported by the European Commission under the Marie Curie Initial Training Network UNILHC 237920 (Unification in the LHC era). Contents
reflect only the authors' views and not the views of the European Commission. We thank Bobby Acharya, Stephen Angus, Erminia Calabrese, Jo Dunkley, Malcom Fairbairn and Andrew Powell for discussions related to this work.

\end{document}